\documentclass[aps,prb,twocolumn,showpacs,a4paper,superscriptaddress]{revtex4-1}
\usepackage{graphicx,bm,amsmath,dcolumn,amssymb}
\graphicspath{{figures}}
\usepackage{geometry,hyperref}
\begin{document}

\title{Reentrance of Berezinskii-Kosterlitz-Thouless-like transitions in three-state Potts antiferromagnetic thin film}
  \author{Chengxiang Ding}
\email{dingcx@ahut.edu.cn}
 \affiliation{Department of Applied Physics, Anhui University of Technology, Maanshan 243002, China }
 \author{Wenan Guo}
\email{waguo@bnu.edu.cn}
 \affiliation{Physics Department, Beijing Normal University, Beijing 100875, China}
 \author{Youjin Deng}
\email{yjdeng@ustc.edu.cn}
 \affiliation{Hefei National Laboratory for Physical Sciences at Microscale, 
 Department of Modern Physics, University of Science and Technology of China, Hefei, 230027, China}

\date{\today} 

\begin{abstract}
Using Monte Carlo simulations and finite-size scaling, 
we study three-state Potts antiferromagnet on layered square lattice 
with two and four layers $L_z=2$ and $4$. 
As temperature decreases, the system develops quasi-long-range order 
via a Berezinskii-Kosterlitz-Thouless transition at finite temperature $T_{c1}$. 
For $L_z=4$, as temperature is further lowered, 
a long-range order breaking the $Z_6$ symmetry develops 
at a second transition at $T_{c2} < T_{c1}$.
The transition at $T_{c2}$ is also Berezinskii-Kosterlitz-Thouless-like, 
but has magnetic critical exponent $\eta=1/9$ instead of the conventional value $\eta = 1/4$.
The emergent $U(1)$ symmetry is clearly demonstrated in the quasi-long-range ordered region 
$T_{c2} \leq T \leq T_{c1}$.
\end{abstract}
\pacs{05.50.+q, 11.10.Kk, 64.60.Cn, 64.60.De}
\maketitle 

\section{Introduction}
\label{intro}
 The $q$-state Potts model\cite{potts, wfypotts} has been studied for long time in statistical physics.
 For the ferromagnetic Potts model, the symmetry of order parameter is simply determined 
 by the Potts spins. The physics is now well understood, thanks to the hypothesis of universality. 
 In contrast, for antiferromagnetic Potts (AFP) model, the order parameter is not only
 associated with the spins but also with the underlying lattice. Thus, one should study case by case.

The properties of the AFP model are often related to extensive ground-state degeneracy, 
which may be caused by frustration\cite{fp} or not\cite{nfp,DengUJ,XiangUJ}. 
The extensive degeneracy of ground states may lead to an ``entropy-driven" finite-temperature phase transition. 
The phase transition is characterized by partial ordered phase at low temperature,
which is ordered on a sublattice of the lattice, satisfying the minimum energy and maximum entropy by 
local modification of the spin states.

The three-state AFP model show good examples of such phase transitions. 
In three-dimensional simple-cubic lattice,
when temperature is high, the model is disordered;
when temperature is sufficiently low, some long-range order develops, 
and the following states could be favored:
the spins on one of the sublattices are ``frozen" at a random Potts value, 
and the spin on any site of the remaining lattice is ``free" to take the other Potts values. 
In Refs. \onlinecite{height} and \onlinecite{Sokalsquare}, such states are called ``ideal" states. 
On the simple cubic lattice, there are six types of such ideal states. 
Thus, the order parameter is of the $Z_6$ symmetry. 
Monte Carlo simulations show that the model undergoes a continuous phase transition\cite{wsk,wang3dprb} between 
the high-temperature disordered phase and the low-temperature partial ordered phase which breaks the $Z_6$ symmetry.
The critical exponents fall into the universality class of three-dimensional XY model\cite{XY3d}.

In two dimensions, the three-state AFP model is extensively studied on different lattices.
On the dice lattice\cite{Sokaldice}, the model undergoes a continuous ordered-disordered
phase transition at finite-temperature.
which belongs to the universality of three-state ferromagnetic Potts model. 
On the honeycomb lattice\cite{honeycomb}, the model is disordered at any temperature, including the zero temperature.
On the kagome lattice\cite{kagome}, the model is disordered at any nonzero temperature but critical at 
zero temperature. The magnetic critical exponent, governing the decay of two-point correlation function,
 is known to be $\eta = 4/3$.
The phase diagram of the model on the square lattice is similar to that on the kagome lattice,
but the critical exponent $\eta=1/3$.  
However, when ferromagnetic next-nearest neighboring (NNN) interactions are included, the model has
two Berezinskii-Kosterlitz-Thouless (BKT) transitions\cite{Nijs}.

 In this work, using a combination of various Monte Carlo algorithms, including the standard Metropolis method, 
 the Wang-Swendsen-Kotek\'{y} (WSK) cluster method \cite{wsk} and the geometric cluster method\cite{gc1,gc2,gc3}, 
 we study the three-state Potts antiferromagnet on the square lattice with multilayers, 
 with antiferromagnetic interactions between layers.
 For the two-layer lattice, we find that the system undergoes 
 a continuous phase transition at finite temperature $T_{c1}>0$. 
 The transition is of BKT type, which has 
 magnetic exponent $\eta=1/4$, different from $\eta=1/3$ for the single-layer system at $T=0$.
 In the whole low-temperature region $0 \leq T \leq T_{c1}$, 
 the $L_z=2$ system is quasi-long-rang ordered, with varying critical exponents $\eta$.
 As the number of layers is increased up to $L_z=4$, 
 we find that beside the BKT transition at $T_{c1}$, the system undergoes a second BKT phase transition at a lower temperature $T_{c2} < T_{c1}$, with critical exponent $\eta=1/9$. 
 When $T< T_{c2}$, a long-range order breaking the $Z_6$ symmetry develops.
 The emergent $U(1)$ symmetry is clearly demonstrated for the quasi-long-range ordered phase $T_{c2} \leq T \leq T_{c1}$.

 The organization of the present paper is as follows. 
 Section. \ref{model} defines the model and the observables to be sampled, 
 and introduces the algorithms used in our simulations. 
 The simulation results, including the results for two-layer and four-layer square lattices, are 
 given in Sec. \ref{resul}. We then finally conclude with a discussion in Sec. \ref{concl}.

\section{Model, Algorithm, and Observable}
\label{model}
The three-state Potts model is defined by a simple Hamiltonian
\begin{eqnarray}
\mathcal{H}=-K\sum\limits_{\langle i,j\rangle}\delta_{\sigma_i,\sigma_j} \label{hamil}
\end{eqnarray}
where the sum takes over all nearest neighboring sites $\langle i,j\rangle$. 
The spin assumes $\sigma_i=1, 2, 3$, and
$K=J/k_BT$ is a dimensionless coupling constant.
The model is ferromagnetic when $J>0$ or antiferromagnetic when $J<0$.
In the current paper, we focus on the antiferromagnetic case and set $J/k_B=-1$
for convenience. 

The Potts spin $\sigma$ can also be written as unit vector in the plane
\begin{eqnarray}
\vec{\sigma} =(\cos\theta,\sin\theta ),
\end{eqnarray}
where $\theta =0,\pm 2\pi/3$ represents the angle of the spin.
The Hamiltonian of the three-state Potts model becomes then
\begin{eqnarray}
\mathcal{H}=-\frac{2}{3}K\sum\limits_{\langle i,j \rangle} \cos(\theta_i-\theta_j) \; ,
\end{eqnarray}
apart from a constant.

For Monte Carlo simulations of the three-state antiferromagnetic Potts model on the single-layer square lattice,
the Wang-Swendsen-Kotek\'{y} (WSK) algorithm \cite{wsk} is efficient even at zero temperature.
On the two-layer lattice, the algorithm still works but the efficiency drops. 
At the low temperatures of the four-layer square lattice, the efficiency drops so much 
that it becomes difficult to give reliable data for systems of moderate sizes. 
To overcome this problem, we implement the geometric cluster algorithm\cite{gc1,gc2,gc3}. 
It is shown that a combination of the geometrical algorithm, the WSK algorithm and the Metropolis algorithm 
significantly improves the efficiency, which enables us to extensively simulate systems with linear size up to $L=512$.

The sampled observables in our Monte Carlo simulations include
the staggered magnetization $m_{\rm s}$, the staggered susceptibility $\chi_s$, 
the uniform magnetization $m_{\rm u}$, the uniform susceptibility $\chi_{\rm u}$,
and the specific heat $C_{\rm v}$, which are defined as
\begin{eqnarray}
m_{\rm s}&=&\langle|\mathcal{M}_{\rm s}|\rangle,\\
\chi_{\rm s}&=&N\langle \mathcal{M}_{\rm s}^2\rangle,\\
m_{\rm u}&=&\langle |\mathcal{M}_{\rm u}|\rangle\\
\chi_{\rm u}&=&N\langle \mathcal{M}_{\rm u}^2\rangle,\\
C_{\rm v}&=&N(\langle \mathcal{E}^2\rangle-\langle \mathcal{E}\rangle^2)/T^2,
\end{eqnarray}
with $\mathcal{M}_{\rm s}$, $\mathcal{M}_{\rm u}$, and $\mathcal{E}$ defined as
\begin{eqnarray}
\mathcal{M}_{\rm s}&=&\frac{1}{N}\sum\limits_{\vec{r}}(-1)^{x+y+z} \vec{\sigma}(\vec{r}),\\
\mathcal{M}_{\rm u}&=&\frac{1}{N}\sum\limits_{\vec{r}}\vec{\sigma}(\vec{r}),\\
\mathcal{E}&=&\frac{1}{N}\sum\limits_{\langle i,j\rangle}\delta_{\sigma_i,\sigma_j}
\end{eqnarray}
where $\vec{r}=(x,y,z)$ is the coordination, $N=L^2\times L_z$ is the number of sites of the lattice.
The staggered magnetization $m_{\rm s}$ can be conveniently used to probe the breaking of the $Z_6$ symmetry in the ordered phase.

We also sample the correlation length $\xi$ on one sublattice of a given layer. 
Specifically, a layered square lattice is divided to two equivalent sublattices according to the 
parity of $x+y+z$, denoted by ``sublattice A" and ``sublattice B"; the $z$-th layer of sublattice 
A is denoted by ${\rm A}_z$. The in-layer sublattice correlation length $\xi$ is then defined as\cite{Sokalsquare}
\begin{eqnarray}
\xi=\frac{(\chi/F-1)^{1/2}}{2\sqrt{\sum\limits_{i=1}^d \sin ^2(\frac{k_i}{2})}} \; ,
\end{eqnarray}
where $\vec{k}$ is the ``smallest wavevector" of the square lattice along the $x$ direction--i.e., $\vec{k} \equiv (2 \pi/L, 0)$.
The in-layer sublattice susceptibility $\chi$ and the ``structure factor" $F$ are 
\begin{eqnarray}
\chi&=&\frac{1}{N}\langle\big|\sum\limits_{\vec{r}{~\rm on~A_z}}\vec{\sigma}(\vec{r})\big|^2 \rangle,\label{chi}\\
F&=&\frac{1}{N}\langle\big|\sum\limits_{\vec{r}{~\rm on~A_z}}e^{i\vec{k} \cdot\vec{r}}
\vec{\sigma}(\vec{r})\big|^2\rangle \; .
\label{F}
\end{eqnarray}
In a critical phase, quantity $\xi/L$ assumes a universal value in the thermodynamic limit $L \rightarrow \infty$.
In a disordered phase, correlation length $\xi$ is finite and $\xi/L$ drops to zero, 
while in an ordered phase, $\xi/L$ diverges quickly since ``structure factor" $F$ vanishes rapidly.
Thus, $\xi/L$ is known to be very useful in locating the critical points of phase transitions.

\section{Results}
\label{resul}
In simulations of the three-state AFP model on multilayer square lattice, periodic boundary condition is used, 
including the $z$ direction. The largest system size in the simulation
is $L=512$ and each data point is averaged over $5\times 10^6 \sim 10^7 $ samples.

\subsection{Three-state AFP model on the two-layer square lattice}
\begin{figure}[htpb]
\includegraphics[scale=0.7]{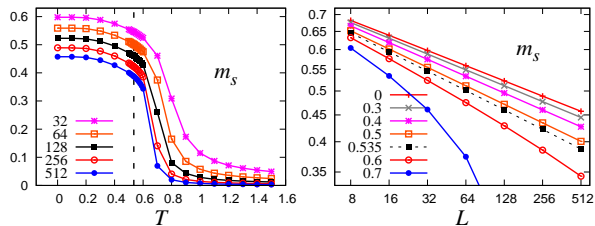}
\caption{(Color online) Left, staggered magnetization $m_s$ versus $T$ for the three-state AFP model
on the two-layer square lattice; the dashed vertical line is set at the critical point $T_{c1}=0.535$.
Right, log-log plot of $m_{\rm s}$ versus $L$ for given temperatures. }
\label{msL2}
\end{figure}

The left of Fig. \ref{msL2} is an illustrative plot of $m_{\rm s}$ versus $T$
 for the two-layer three-state AFP model for a series of system sizes.
The figure shows that in high temperature, the staggered 
magnetization converges to zero; in low temperature, the magnetization also decreases
as the system size increases; however the finite-size scaling behavior in this region is obviously different
to that in the high-temperature region. This is shown more clearly by the log-log plot of $m_{\rm s}$ versus $L$
for given temperatures, as shown in the right of Fig. \ref{msL2}.
We find that the magnetization $m_{\rm s}$ in the low temperatures can be described by 
\begin{eqnarray}
  m_{\rm s}&=& L^{y_{\rm s}-d}(a+b_1/\ln L+b_2L^{y_i}), \label{msfss}
\end{eqnarray}
where $d=2$ is the spatial dimension, and $y_{\rm s}$ is renormalization exponent of the staggered magnetic field,
which varies continuously with the temperature.
$b_1/\ln L$ and $b_2L^{y_i}$ are the correction-to scaling terms, with $y_i<0$. $a, b_1$, and $b_2$ are unknown parameters.

The meaning of the scaling behavior of $m_{\rm s}$ in the low-temperature region is twofold: 
First, it means that the staggered magnetization in the 
low-temperature region also converges to zero as the system size increases to infinite (because $y_{\rm s}<2$).
This means that in the thermodynamic limit the $Z_6$ symmetry in the system 
is not broken, namely the system doesn't have long-range order on the sublattices.
Second, it implies that this region is critical and the phase transition 
is of the BKT type.
This result is confirmed by the behaviors of the staggered susceptibility.
In the low temperatures, the staggered susceptibility scales as
\begin{eqnarray}
  \chi_{\rm s}&=& L^{2y_{\rm s}-d}(a+b_1/\ln L+b_2L^{y_i}). \label{chisfss}
\end{eqnarray}
The values of $y_{\rm s}$ at different temperatures can be obtained by fitting 
(\ref{msfss}) or (\ref{chisfss}) to the data; the best estimations are listed in Table \ref{L2tab}.

 \begin{figure}[htpb]
 \includegraphics[scale=0.7]{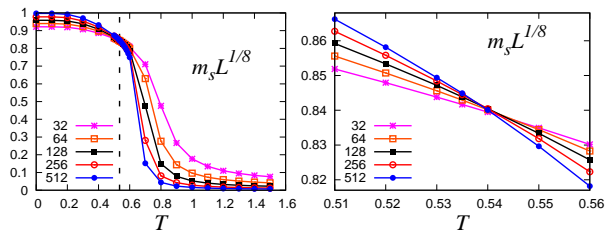}
 \caption{(Color online) Left, plot of $m_sL^{1/8}$ versus $T$ for various $L$.
 Right, an enlarge version of the left plot nearing the critical point.
 The dashed vertical line is set at the critical point $T_{c1}=0.535$.}
 \label{mscriL2}
 \end{figure}

The critical point can be located more accurately by the finite-size scaling behavior 
of $m_{\rm s}$. Figure \ref{mscriL2} is a plot of $m_sL^{d-y_{\rm s}}$ versus $T$, where we have 
set the value of $y_{\rm s}$ as the exact one for BKT transition, i.e., $y_{\rm s}=15/8$.
It obviously indicates a transition at $T_{c1}\approx 0.54$. Fitting the data nearing 
this point by the following formula
\begin{eqnarray}
m_{\rm s}&=&L^{y_{\rm s}-d}[a_0+\sum\limits_{k=1}^{2}a_k(T-T_{c1})^k(\ln L)^k\nonumber\\
&&+\sum\limits_{j=1}^2b_j(T-T_{c1})^j+\frac{c_0}{\ln L}+c_1L^{y_{_i}}],\label{mscrifss}
\end{eqnarray}
we get $T_{c1}=0.535(3)$.
At this point, $y_{\rm s}$ is estimated to be 1.875(1), which coincides with the exact result 15/8.
This gives a self-consistent check.

\begin{figure}[htpb]
\includegraphics[scale=0.82]{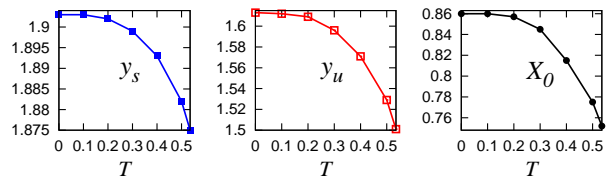}
\caption{(Color online) Critical exponents $y_{\rm s}$, $y_{\rm u}$ and $X_0$ versus $T$ for the 
three-state AFP model on the two-layer square lattice.}
\label{expL2}
\end{figure}

The uniform magnetization $m_u$ and uniform susceptibility $\chi_{\rm u}$ are also calculated in the simulations.
It is found that $m_u$ and $\chi_{\rm u}$ show similar scaling behaviors as $m_{\rm s}$ and $\chi_{\rm s}$ 
respectively,
with the staggered exponent $y_{\rm s}$ replaced by a uniform exponent $y_{\rm u}$. 
Doing similar fitting, $y_{\rm u}$ are obtained, which are also listed in Table \ref{L2tab}.
Figure \ref{expL2} is an illustrative plot of the critical exponents versus temperature $T$.

\begin{figure}[htpb]
\includegraphics[scale=0.7]{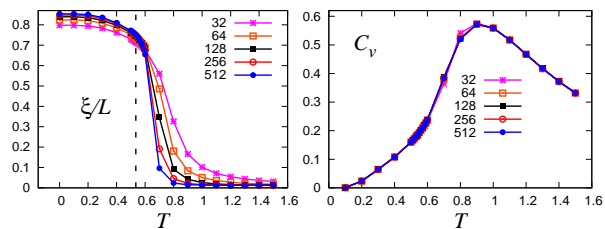}
\caption{(Color online) Plot of $\xi/L$ and $C_{\rm v}$ versus $T$ for various system sizes for the 
three-state AFP model on the two-layer square lattice; the dashed vertical line is set at the BKT point $T_{c1}=0.535$.}
\label{xi}
\end{figure}

The BKT transition can also be demonstrated by the behavior of $\xi/L$, as shown in 
the left of Fig. \ref{xi}.
In the region $T>T_{c1}$, the value of $\xi/L$ converges to zero as the system size increases;
in the region $T\le T_{c1}$, the value of $\xi/L$ converges to a finite nonzero value,
which can be fit according to 
\begin{eqnarray}
\xi/L=X_0+b_1/\ln L+b_2L^{y_i}+\cdots,\label{xifss}
\end{eqnarray}
with $y_i<0$.
Table. \ref{L2tab} lists the results of $X_0$ for different temperatures. 

In fitting the data according to Eqs. (\ref{msfss}), (\ref{chisfss}), (\ref{mscrifss}), and (\ref{xifss}), 
the logarithmic terms are included. In fact, the BKT transition is characterized 
by logarithmic corrections\cite{logxy,logxy2,logxy3,logxy4}, 
due to the presence of marginally relevant temperature field in renormalization\cite{planar}.

\begin{table}[htbp]
\caption{Critical exponents of the three-state antiferromagnetic Potts model 
on the two-layer square lattice.}
 \begin{tabular}{l l l l}
    \hline
    \hline
      $T$   &$y_{\rm s}$&$y_{_{\rm u}}$&$X_0$\\
    \hline
    $0.0$   &1.903(2) &1.613(2)&0.860(5)\\
    $0.1$   &1.903(2) &1.612(2)&0.860(5)\\
    $0.2$   &1.902(2) &1.609(2)&0.857(5)\\
    $0.3$   &1.899(2) &1.596(2)&0.845(5)\\
    $0.4$   &1.893(2) &1.571(2)&0.815(5)\\
    $0.5$   &1.882(2) &1.529(2)&0.775(5)\\
    $0.535$ &1.875(3) &1.501(3)&0.752(5)\\
    \hline
    \hline
\end{tabular}
\label{L2tab}
\end{table}

At last, we present the result for the specific heat of the model, as shown in the right of Fig. \ref{xi}. It is seen that
the specific heat doesn't diverge but has a broad peak which converges to finite value.
This is also the typical character of BKT transition.

\subsection{The three-state AFP model on four-layer square lattice}
\begin{figure}[tbhp]
\includegraphics[scale=0.7]{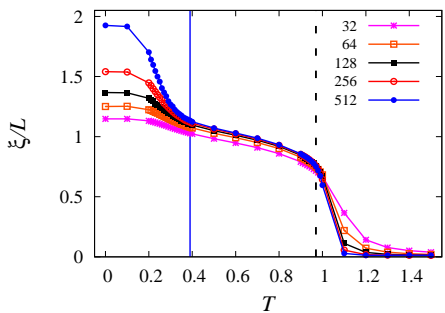}
\caption{(Color online) $\xi/L$ versus $T$ for the three-state AFP model on the four-layer square lattice.
 The dashed vertical line is
set at $T_{c1}=0.97$; the solid vertical line is set at $T_{c2}=0.39$.}
\label{xiL4}
\end{figure}

On the four-layer square lattice, the three-state AFP model undergoes two BKT-like transitions,
which can be clearly demonstrated by the critical behavior of $\xi/L$, as shown in Fig. \ref{xiL4}. 
At high temperature, the system is disordered and the value of $\xi/L$ converges to zero as system size
$L \to \infty$; at low temperature, the system is ordered which breaks the $Z_6$ symmetry and the 
value of $\xi/L$ diverges; 
at the intermediate temperatures, 
the system is quasi-long-range ordered and the value of 
$\xi/L$ converges to finite nonzero value $X_0$. 
By fitting the data according to (\ref{xifss}), a series of $X_0$ are obtained and listed in 
Table \ref{L4tab}.

\begin{table}[htbp]
\caption{Critical exponents of the three-state antiferromagnetic Potts model 
on the four-layer square lattice.}
 \begin{tabular}{llll}
    \hline
    \hline
      $T$   &$y_{\rm s}$&$y_{\rm u}$&$X_0$\\
    \hline
    $0.39$  &1.944(2) &1.777(2)&1.15(1)\\
    $0.5$   &1.938(2) &1.752(2)&1.09(1)\\
    $0.6$   &1.933(2) &1.731(2)&1.04(1)\\
    $0.7$   &1.926(2) &1.704(2)&0.99(1)\\
    $0.8$   &1.917(2) &1.669(2)&0.93(1)\\
    $0.9$   &1.903(2) &1.612(2)&0.86(1)\\
    $0.97$  &1.874(3) &1.501(3)&0.75(1) \\
    \hline
    \hline
\end{tabular}
\label{L4tab}
\end{table}

\begin{figure}[htpb]
\includegraphics[scale=0.7]{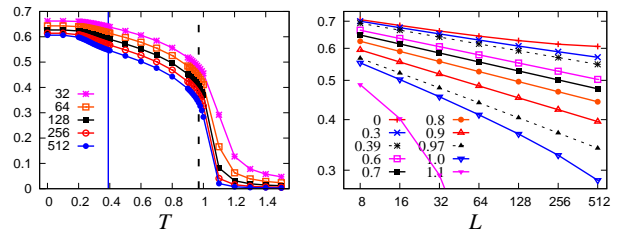}
\caption{(Color online) Left, $m_s$ versus $T$ for the three-state AFP model on 
the four-layer square lattice; the dashed vertical line is set at the BKT point $T_{c1}=0.97$, the solid vertical 
line is set at the BKT point $T_{c2}=0.39$.
Right, log-log plot of $m_s$ versus $L$ for the three-state AFP model on the four-layer square lattice; the two dashed 
lines correspond to the BKT points.}
\label{msL4}
\end{figure}

The two BKT-like transitions are further illustrated by the behavior of $m_s$ in the left of 
Fig. \ref{msL4}. At  high temperatures, the magnetization converges to zero as the system
size increases; at  low temperatures, it converges to nonzero value which
indicates the break of $Z_6$ symmetry; in the intermediate temperatures, it scales as (\ref{msfss}). 
The right of Fig. \ref{msL4} is a log-log plot of $m_s$ versus $L$ for given temperatures, 
which shows the finite-size scaling behavior of $m_s$ more clearly.
The values of the critical exponent $y_{\rm s} $ in the quasi-LRO phase, 
obtained by fitting the data according to (\ref{msfss}),
are also listed in Table \ref{L4tab}.
The fit is perfect in the region $0.39\le T\le 0.97$ but deteriorates when $T<0.97$ or $T>0.39$;
this implies the critical points $T_{c1}\approx 0.97$ and $T_{c2}\approx 0.39$. 

\begin{figure}[htpb]
\includegraphics[scale=0.82]{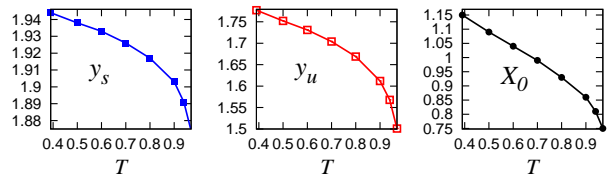}
\caption{(Color online) Critical exponents $y_{\rm s}$, $y_{\rm u}$ and $X_0$ versus $T$ for the 
three-state AFP model on the four-layer square lattice.}
\label{expL4}
\end{figure}

Similar critical behaviors are observed for $m_u$ and $\chi_{\rm u}$, the estimated values of $y_{\rm u}$ 
are listed in Table \ref{L4tab}.
Figure \ref{expL4} is an illustrative plot of the critical exponents versus temperature $T$.

The curve of the specific heat of the four-layer model is very similar to that of the two-layer model (right of Fig. \ref{xi});
it has one and only one finite peak (but not diverge) at $T_{c1}$; it does not show any singularity. 

\begin{figure}[htpb]
\includegraphics[scale=0.7]{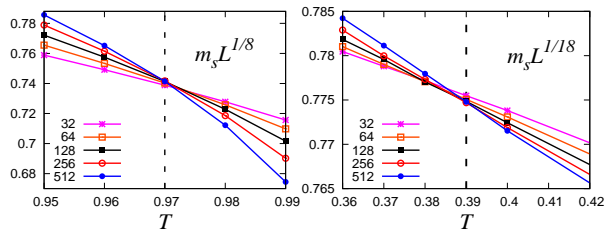}
\caption{(Color online) Plot of $m_sL^{y_{\rm s}-d}$ versus $T$ for the three-state AFP model on 
the four-layer square lattice. Left, $y_{\rm s}=15/8$; right, $y_{\rm s}=35/18$.}
\label{mscriL4}
\end{figure}

The phase diagram of the three-state AFP model on the four-layer lattice is similar to that on the 
single-layer square lattice with ferromagnetic NNN interactions\cite{Nijs}. 
The latter can be mapped onto a Gaussian model, and the critical exponents $y_{\rm s}$ and 
$y_{\rm u}$ are determined by the vortex excitations in the Gaussian model with charge 
$\pm 1$ and $\pm 2$ respectively, with
\begin{eqnarray}
y=2-\frac{n^2}{4\pi K_G}. \label{yy}
\end{eqnarray}
Here $y=y_{\rm s}$ or $y_{\rm u}$, $n$ is the charge;
$K_G$ is the coupling constant of the Gaussian model.
For $T=T_{c1}$, $K_G=2/\pi$, thus $y_{\rm s}=15/8$ and $y_{\rm u}=3/2$; for $T_{c2}$, $K_G=9/2\pi$, 
thus $y_{\rm s}=35/18$ and $y_{\rm u}=16/9$.
Assuming these results are also valid for the three-state AFP model on the multilayer lattice,
we plot $m_sL^{d-y_{\rm s}}$ versus $T$ with $y_{\rm s}=15/8$ in the left of Fig. \ref{mscriL4},
which obviously shows the phase transition at $T_{c1}$. Fitting the data according to (\ref{mscrifss}) with $y_{\rm s}=15/8$ fixed, 
we obtained the critical point $T_{c1}=0.967(5)$.
The right of Fig. \ref{mscriL4} is a plot of $m_sL^{d-y_{\rm s}}$ versus $T$ with $y_{\rm s}=35/18$, 
which obviously shows the phase transition at $T_{c2}$. A Similar fit yields $T_{c2}=0.393(5)$.

Furthermore, from Tables \ref{L2tab} and \ref{L4tab} we find that $y_{\rm s}$ and $y_{\rm u}$ satisfy 
\begin{eqnarray}
2-y_{\rm u}=4(2-y_{\rm s}).
\end{eqnarray}
This is also the case for the three-state AFP model on the single-layer square lattice with 
ferromagnetic NNN interactions, which can be easily derived from Eq. (\ref{yy}).

We also calculate the observables concerning the rotational symmetry of the model
\begin{eqnarray}
  \phi_6&=&\cos 6\theta\\
Q_{\phi}&=&\frac{\langle\phi_6^4\rangle}{\langle\phi_6^2\rangle^2},
\end{eqnarray} 
where $\theta$ is defined as the angle of the vector $\mathcal{M}_s$
\begin{eqnarray}
\theta=
\left\{
\begin{array}{cc}
 \tan^{-1}(\mathcal{M}_y/\mathcal{M}_x)+\pi/2, & {\rm if~}\mathcal{M}_x>0\\
 \tan^{-1}(\mathcal{M}_y/\mathcal{M}_x)+3\pi/2, & {\rm ~if~}\mathcal{M}_x<0.
\end{array}
\right.
\end{eqnarray} 
Here $\mathcal{M}_x,\mathcal{M}_y$ are the two components of $\mathcal{M}_s$. This definition makes the value of 
$\theta$ be in the region $[0,2\pi]$. $Q_{\phi}$ is known to be useful in distinguishing the quasi-LRO phase
and the true LRO phase\cite{gencl}.
\begin{figure}[tbhp]
\includegraphics[scale=0.7]{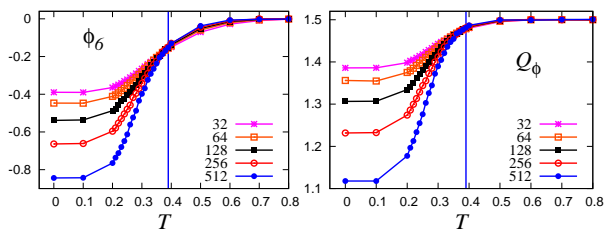}
\caption{(Color online) Plot of $\phi_6$ and the Binder ratio $Q_\phi$ for the three-state 
AFP  model on the four-layer square lattice. The vertical lines are set at the critical point $T_{c2}=0.39$.}
\label{fi}
\end{figure}
\begin{figure}[htpb]
\includegraphics[scale=0.40]{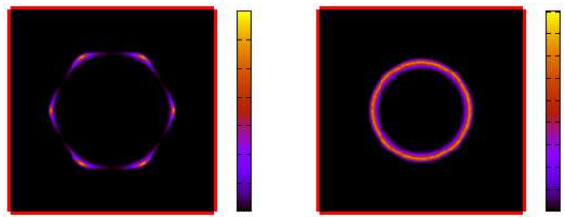}
\caption{(Color online) Histogram of ($\mathcal{M}_x,\mathcal{M}_y$) for the three-state AFP model on the four-layer square lattice,
 with system size $L=512$. Left, $T=0.2$; right, $T=0.7$.}
\label{hist}
\end{figure}

Figure \ref{fi} is the plot of $\phi_6$ and $Q_\phi$; it obviously indicates a phase transition 
at $T_{c2}\approx0.39$.
Figure \ref{hist} is the plot of the histogram of ($\mathcal{M}_x$, $\mathcal{M}_y$). 
These results are easy to understand. In the the quasi-LRO phase $m_s$ is zero in the thermodynamic limit, 
thus the angle of $\mathcal{M}_s$ can take random value in $[0,2\pi]$; it is consistent with 
the emergent $U(1)$ symmetry of $\mathcal{M}_s$ in a finite system.
In the low-temperature phase, $m_s$ is not zero and the $Z_6$ symmetry is broken, the 
angle of $\mathcal{M}_s$ favors six directions in a finite system, as shown in the left of 
Fig. \ref{hist}. 

\section{Conclusion and discussion}
\label{concl}
In conclusion, we have studied the three-state antiferromagnetic Potts model on the layered
square lattice. On the two-layer lattice, the model undergoes a BKT-like transition
at $T_{c1}=0.535(3)$, with critical exponent $y_{\rm s}=15/8$ ($\eta=1/4$).
On the four-layer lattice, the model has two BKT-like transitions.
One is between the high-temperature phase and the quasi-long-range ordered phase,
with critical point $T_{c1}=0.967(5)$ and critical exponent $y_{\rm s}=15/8$ ($\eta=1/4$). 
Another one is between the quasi-long-range ordered phase and the low-temperature ordered phase which 
breaks the $Z_6$ symmetry, with critical point $T_{c2}=0.393(5)$ and critical exponent $y_{\rm s}=35/18$ ($\eta=1/9$).
Emergent $U(1)$ symmetry is found in the quasi-long-range ordered phase.

The critical properties of the three-state AFP model on the square lattice 
are related to the vortex excitations, which is investigated by Kolafa\cite{afpvortex} using Monte Carlo simulations. 
The simulations show that the positive and negative vortices are bound into dipoles only at 
the zero temperature; at any temperature $T>0$, the dipoles unbind. 
There is no quasi-LRO phase in the single-layer AFP model.
However, our simulations show that the multilayer structure can lead to a quasi-LRO phase.
This is due to the modifications of the ground states by the layered structure.
On a bipartite lattice such as the square lattice, the simple cubic lattice, or 
the layered square lattice, the density of entropy of the ideal states is $s_i=\ln2/2=0.3466$. On the square lattice,
the density of total entropy of the model is $s=3/2\ln(4/3)$\cite{lieb}. The ratio is $r_s=s_i/s=0.803$. 
On the simple cubic lattice $s=0.367$ and $r_s=0.945$. For the layered square lattice, although we 
have not numerically calculated its entropy density, we believe it
is reasonable to postulate
that the value of $r_s$ is between 0.803 and 0.945; and it will gradually increase as the number of 
layers increases. 
The increase of $r_s$ means the enhancement of the effect of ideal states,
which will restrict the vortex excitations with nonzero charge, because
the vortex excitations based on ideal states favor zero charge.
As pointed out by Ref. \onlinecite{Nijs}, the zero charge does not 
dominate the leading critical properties of the model.
Therefore, comparing to the single-layer model, the multilayer model needs higher temperature to 
generate vortices with
nonzero charge, thus the quasi-LRO phase of the two-layer model enters the region with $T>0$. 
Another obvious result of the increase of $r_s$ is the enhancement of the effect of $Z_6$ symmetry,
which tends to make the system be ordered. However when the number of layers is two, the effect is not 
strong enough. When the number of layers increases to four, it is strong enough to lead to 
an ordered phase. 

The phase diagram of the four-layer square-lattice AFP model is very similar to that of the single-layer square-lattice 
AFP model with ferromagnetic NNN interactions\cite{Nijs}. 
For the latter, the ferromagnetic NNN interactions have similar effect as that of the multilayer structure; it  also
enhances the $Z_6$ symmetry and restricts the vortex excitations with nonzero charge.
However, the two model still have some subtle difference.
For the single-layer model, if the ratio of the strength of the NNN interactions 
and the NN interactions takes fixed nonzero value, the system must be ordered at zero temperature.
In such a case, it does not have a single BKT transition like that in the two-layer lattice.
Furthermore, the entropy of the ground states of the single-layer model is not extensive; this is obviously different to
that of the multilayer lattice, although it doesn't lead to substantial difference in critical behaviors.

For ferromagnetic model on multilayered lattice, the phase transition behavior belongs to the same
universality class as that in the corresponding single-layer lattice\cite{fpl,tli2,tli}, according to the hypothesis of universality. 
However, for antiferromagnetic model, due to the lattice structure dependence nature of the model,
the number of layers may lead to substantially different behavior of phase transition from that in the corresponding single-layer lattice.

\section{Acknowledgment}
This work is supported by the National Science Foundation of China
 (NSFC) under Grant Nos. 11205005 (Ding), 11175018 (Guo), and 11275185 (Deng).

\end{document}